\documentclass[a4paper, 10pt]{article}
\usepackage{authblk}
\usepackage[authoryear]{natbib}
\usepackage[left=1in, right=1in, top=0.3in, bottom=0.7in]{geometry}

\begin{document}
\renewcommand{\thefootnote}{\arabic{footnote}}

\title{\huge Second data release of the {\it Kepler}-INT Survey}
\author{}
\date{}
\maketitle
\vspace*{-1.5cm}
\begin{center}
\author{S. Greiss\footnote{Department of Physics, Astronomy and Astrophysics group, University of Warwick, CV4 7AL, Coventry, U.K.}, D. T. H. Steeghs$^{1}$, B. T. G\"ansicke$^{1}$, E. L. Mart\'in\footnote{INTA-CSIC Centro de Astrobiolog\'{i}a, Carretera de Ajalvir km 4, 28550 Torrej\'{o}n de Ardoz, Spain}, P. J. Groot\footnote{Department of Astrophysics/IMAPP, Radboud University Nijmegen, P.O.Box 9010, 6500 GL, Nijmegen, The Netherlands}, M. J. Irwin\footnote{Cambridge Astronomy Survey Unit, Institute of Astronomy, University of Cambridge, Madingley Road, CB3 0HA, Cambridge, U.K.}, E.~Gonz\'alez-Solares$^{4}$, R. Greimel\footnote{Institut f\"ur Physik, Karl-Franzen Universit\"at Graz, Universit\"atsplatz 5, 8010 Graz, Austria}, N. P. Gentile Fusillo$^{1}$, M. Still\footnote{NASA Ames Research Center, M/S 244-40, Moffett Field, CA 94035, USA} \footnote{Bay Area Environmental Research Institute, Inc., 560 Third St. West, Sonoma, CA 95476, USA}, the KIS collaboration}\\
\vspace*{0.5cm}
s.greiss@warwick.ac.uk

\end{center}
\section*{Report}
This short document concerns the second data release of the {\it Kepler}-INT Survey (KIS, \citealt{greissetal12}). The {\it Kepler} field, a 116 deg$^{2}$ region of the Cygnus and Lyra constellations, is the target of the most intensive search for transiting planets to date. The {\it Kepler} mission provides superior time series photometry, with an enormous impact on all areas of stellar variability. However, up to the beginning of 2012, its field lacked optical photometry complete to the confusion limit of the Kepler instrument. For this reason, in 2011, we followed the observing strategy and data reduction method used in the IPHAS \citep{drewetal05} and UVEX \citep{grootetal09} galactic plane surveys in order to produce a deep optical survey of the {\it Kepler} field. The initial release catalogue \citep{greissetal12} concerned data taken between May and August  2011, using the Isaac Newton Telescope on the island of La Palma. Four broadband filters were used, $U, g, r, i$, as well as one narrowband one, H$\alpha$, reaching down to a limit of $\sim$20$^{th}$ mag in the Vega system. Observations covering $\sim$50 deg$^{2}$ passed our quality control thresholds and constituted the first data release. Here we report on the second data release, covering an additional $\sim$63 deg$^2$ of the {\it Kepler} field. We apply the exact same quality control criteria and photometric calibration as in the initial release catalogue (see \citet{greissetal12} for further details). The global photometric calibration was derived by placing the KIS magnitudes as close as possible to the {\it Kepler} Input Catalog (KIC) ones. The second data release catalogue containing $\sim$14.5 million sources from all the good photometric KIS fields is available for download from the KIS webpage ($\mathrm{www.astro.warwick.ac.uk/research/kis/}$), as well as from the {\it Kepler} Target Search page at MAST ($\mathrm{http://archdev.stsci.edu/kepler/kepler\_fov/search.php}$). It includes all the objects from the initial data release catalogue and covers $\sim$97\% of the {\it Kepler} field. Therefore this release adds $\sim$8.5 million sources to the initial data release catalogue, all observed between May and August 2012. The remaining $\sim$3\% of the field will be completed once the {\it Kepler} field is visible in 2013. Similarly to the initial data release catalogue, the number of sources given does not correspond to the number of unique sources in the catalogue. Around 25\% of the sources will have duplicates and a few thousand sources have more than 2 detections in the catalogue.

\section*{Acknowledgments}
This paper makes use of data collected at the Isaac
Newton Telescope, operated on the island of La Palma,
by the Isaac Newton Group in the Spanish Observatorio
del Roque de los Muchachos. The observations were
processed by the Cambridge Astronomy Survey Unit
(CASU) at the Institute of Astronomy, University of
Cambridge. We acknowledge the use of data taken from
the Kepler Input Catalog (KIC), as well as data taken
from the Sloan Digital Sky Survey.

\noindent S. Greiss acknowledges support through the Warwick Postgraduate Research Scholarship. D. Steeghs acknowledges a STFC Advanced Fellowship.

\end{document}